\documentclass[11pt,letterpaper]{article}
\pdfoutput=1
\usepackage{jheppub}
\usepackage{natbib}
\usepackage{wrapfig}

\usepackage{tikz}
\usetikzlibrary{calc}

\usepackage{hyperref}
\usepackage{amsmath}
\usepackage{amsfonts}
\usepackage{graphicx}
\usepackage{caption}
\usepackage{placeins}
\usepackage{tqft}
\usepackage{comment}
\usepackage{array}

\DeclareRobustCommand{\C}{\mathcal{C}}

\title{1-loop Amplitudes from the Halohedron}

\author[a,c]{Giulio Salvatori}

\affiliation[a]{Dipartimento di Fisica, Universit\`a degli Studi di Milano,
Via Celoria 16, IT-20133 Milano, Italy}
\affiliation[c]{INFN, Sezione di Milano, \\
Via Celoria 16, I-20133 Milano, Italy.}

\emailAdd{giulio.salvatori@unimi.it}

\abstract{We recently proposed the Halohedron to be the 1-loop Amplituhedron for planar $\phi^3$ theory. Here we prove this claim by showing how it is possible to extract the integrand for the partial amplitude $m^1_n(1,\dots,n|1,\dots,n)$ from the canonical form of an Halohedron which lives in an abstract space. This space is just a step away from ordinary kinematical space at 1-loop, because it is composed by abstract variables associated to propagators of 1-loop Feynman diagrams. Such variables, however, are unbound from momentum conservation relations that would give problems such as double poles.
As an application of our construction, we exploit a well known recursion formula for the canonical form of a polytope in order to produce an expression for the 1-loop integrand which would not be evident starting from Feynman diagrams.}

\begin{document}
\maketitle
\flushbottom

\section{Introduction}

In a recent work we proposed the Halohedron to be the 1-loop Amplituhedron for the planar $\phi^3$ theory \cite{Salvatori:2018fjp}.
The Halohedron emerged naturally by using hyperbolic geometry in the study of positive geometries living in the moduli space of genus one Riemann surfaces, $\mathcal{M}_{1,n}$.
Such considerations were most natural in (1+2) dimensions\footnote{For another approach where 1+2 dimensions were instrumental, see \cite{Cachazo:2013iaa}}, where the hyperboloid model provides a simple way to solve the \emph{scattering equations}\cite{Cachazo:2013gna,Cachazo:2013hca,Cachazo:2013iea}, which, at least at tree level, are known to provide maps from positive geometries defined in kinematical space to positive geometries in the moduli space \cite{Cachazo:2016ror,Arkani-Hamed:2017mur}.

However, choosing a specific dimension clashes with the general wisdom of the scattering equations, which is instead working in \emph{arbitrary} dimension.
Indeed, in our case it proved to be a substantial obstacle in extracting the amplitude from the canonical form of these positive geometries, even at tree level. This is mainly due to the fact that, if a specific dimension is picked, then the Mandelstam variables have to satisfy non-linear Gram identities that do not interact well with the constraints used to cut the Associahedron. An interesting approach to this problem may be to think of (1+2)-dimensional kinematical space, or any other d-dimensional kinematical space, as a particular subspace of the arbitrary dimension Mandelstam space, obtained from the latter by imposing the Gram identities. This idea is inspired by \cite{He:2018pue}, where it was shown how to extract amplitudes from general subspaces of the  Mandelstam space, and surely deserves further study.

\begin{figure}[!htbp]
\begin{center}

\begin{tikzpicture}[>=latex,scale=4]
\foreach \x/\y in {-0.8368/0.3636,-0.9218/0.267,-0.9721/0.1317,-0.9218/-0.1775,-0.8098/-0.3128}
{ \draw[thick] (\x,\y) -- (-0.5968,0);}

\foreach \x/\y in {0.7247/-0.3244,0.8368/-0.2316,0.9103/-0.05771,0.9141/0.1394,0.8136/0.3559}
{ \draw[thick] (\x,\y) -- (0.5968,0);}

\draw[thick] (-0.5968,0.) -- (0.5968,0.) node(h)[pos = 0.25,above]{$s_I$} node(i)[pos = 0.75,above]{$s_I$};

\filldraw[gray] (-0.5968,0.) circle (0.2);
\filldraw[gray] (0.5968,0.) circle (0.2);

\filldraw[white] (0,0.) circle (0.2);
\draw[thick] (0,0.) circle (0.2);

\end{tikzpicture}

\caption{The topology of bubble diagrams make so that double poles arise}
\label{fig:diagrammabolla}

\end{center}
\end{figure}
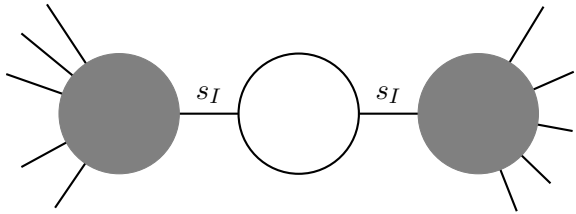

As one might expect, at 1-loop there are new sources of problems. Whilst tadpoles and external leg bubbles (in dimension $D >6$) are known to cancel \cite{He:2015yua},  internal bubbles have some issues which we would like to explain from different perspectives. Firstly, internal bubbles cause the integrand - however defined - to have double poles.
Indeed, a diagram such as in Fig. \ref{fig:diagrammabolla} gives a contribution to the integrand of the form\footnote{We stick to the convention that, for any subset $I$ of the external particle labels $\{1,\dots,n\}$,  $s_I$ denotes a multi-particle Mandelstam invariant, i.e. $s_I = k_I^2$ and $k_I^\mu = \sum\limits_{i \in I} k_i^\mu$ where $k_i$ are the external momenta.} $1/s_I^2$. One may try to get around this, by exploiting the fact that a shift in the loop variable allows to rewrite the internal 2-point function as $s_I/(\ell\cdot k_I)^2$, but then a new double pole $\ell \cdot k_I$ arises. It is difficult to understand how an integrand with double poles may emerge from the canonical form of a positive geometry, which is defined to have \emph{simple} poles\footnote{We would like to thank Nima Arkani-Hamed for stressing this point to us.}! Another reason why bubbles are a problem is that we would like to interpret the propagators of a Feynman diagram as \emph{coordinates} over a positive geometry which should be $n$-dimensional (the dimension of $\mathcal{M}_{1,n}$). The Halohedron is the natural candidate, but then again we do not know how to treat bubbles, that give only $n-1$ independent propagators. In this sense, external leg bubbles and tadpoles are a problem as well, because momentum conservation force a propagator to be identically 0!

In this paper we propose a very simple way to overcome all these problems. The key idea, quite similar to the Big Kinematic Space proposed in \cite{Arkani-Hamed:2017mur}, is to loosen the propagators from the constraints coming from momentum conservation, which forces tadpoles and external bubbles propagators to be zero, and internal bubble propagators to be equal. Therefore, we think of the propagators as abstract variables $X_I$ and we cut an Halohedron in this space. 

Another crucial step is noting that two vertices of the Halohedron lie on the same 1-dimensional edge if the corresponding Feynman diagrams are related by a simple generalisation of the \emph{mutation} introduced in \cite{Arkani-Hamed:2017mur}. This basic observation has the consequence that the canonical form of the Halohedron, once a reference diagram $g^*$ has been fixed, can be written as a sum over \emph{all} 1-loop planar diagrams
\begin{align*}
\Omega_{\mathrm{H}_n} = \pm d\mu_{g^*} \left(\sum_{g} \frac{1}{\prod\limits_{I \in g} X_I} \right)
\end{align*}
where the product $\prod\limits_{I \in g}$ runs over all propagators of the diagram $g$ and the measure $d\mu_{g^*}$, given by $d\mu_{g^*} = \bigwedge\limits_{I \in g^*} dX_I$, is defined up to an overall sign.

To obtain the 1-loop integrand for bi-adjoint theory from $\Omega_{\mathrm{H}_n}$, we strip off the measure $d\mu_{g^{*}}$, kill the tadpoles and external bubbles contribution by sending the corresponding variables to infinity and finally go back to the physical kinematical space by substituting $X_I \to s_I$. Note that as a consequence of these replacements momentum conservation and double poles are restored. In this sense, we can finally state that the Halohedron is the 1-loop Amplituhedron for planar $\phi^3$ theory. 

The new geometrical picture allows to find new recursion formulae for the 1-loop amplitudes by exploiting the standard machinery developed in \cite{Arkani-Hamed:2017mur,Arkani-Hamed:2017tmz}. For example, for the 4-point 1-loop integrand\footnote{This is the integrand associated to a single color order. The corresponding amplitude is $m^1_n(1,\dots,n|1,\dots,n)$ following the notation of \cite{He:2015yua}} we find the expression
\begin{align}
\label{eq:newformula}
\mathcal{I}_4 &= -\frac{1}{\ell_1^2 \ell_2^2 \ell_3^2 \ell_4^2} + \left(\frac{(\ell_1^2 + s_{12}) (\ell_3^2 + s_{12}  )}{s^2_{12} \ell_2^2 \ell_3^2  \ell_4^2  (\ell_3^2 - \ell_1^2)} + \mathrm{cyclical}\right),
\end{align}
where $\ell_i$ is the momentum flowing through particles $i$ and $i+1$. The individual terms of this expansion cannot be obtained recombining Feynman diagrams. In addition, they possess spurious poles which cancel only in the cyclical sum, as it is usual the case when we triangulate a positive geometry.

This paper is structured as follows. In Section \ref{The Abstract space Halohedron} we describe the convex realisation of the Halohedron in an abstract kinematical space and in Section \ref{The Canonical form and the Integrand} we show how to extract the 1-loop integrand for bi-adjoint theory from its canonical form. In Section \ref{Recursion Formula for the 1-loop Integrand} we illustrate how to obtain recursive formulae for the integrand by means of a simple triangulation of the Halohedron, in particular we provide a detailed derivation of \eqref{eq:newformula}. We conclude by discussing some directions for future investigations.

\section{The Abstract space Halohedron}
\label{The Abstract space Halohedron}

In this section we define a convex realisation of the Halohedron in an affine space $X$ with coordinates $(X_1, \dots, X_n)$.  We are going to do so by defining a set of linear functions $X_I$ such that the region where they are all positive cuts an Halohedron. These functions will be in 1-1 correspondence with the facets of the Halohedron, and thus with propagators of 1-loop planar diagrams\footnote{For further details on the correspondence between facets of the Halohedron, arcs on the annulus and propagators of Feynman diagrams see \cite{Salvatori:2018fjp}.}, and will be labelled as the planar variables listed in Tab. \ref{tab:facethaloedron}. The Halohedron is then realised as the intersection of the region where all loop propagators are positive with the space $X$. We can think of the space $X$ as an abstraction of the natural kinematical space of all planar variables, it is a subspace where the planar variables satisfy relations that guarantee the realisation of the Halohedron, while they do not satisfy other usual relations. For example, momentum conservation is not enforced since the planar variables $X_{(i, \dots, j)}$ and $X_{(j+1,\dots,i-1)}$ - which are dual to the propagators at the sides of an internal bubble - will not be equal on $X$. Indeed, this would not be possible since they correspond to different facets of the Halohedron.

\begin{table}
\centering
\resizebox{\textwidth}{0.3\textwidth}{
\begin{tabular}{|c|c|c|}
\hline
Geodesical Arc & Facet type & Planar Variable \\
\hline
\parbox[c]{0.4\linewidth}{
\begin{tikzpicture}[>=latex,scale=0.5, every node/.style={transform shape}]
\filldraw [cyan!30!white] (0,0) circle (2.4cm);
\filldraw [white] (0,0) circle (0.8cm);
\draw [ultra thick] (0,0) circle (2.4cm);
\draw [ultra thick,red] (0,0) circle (0.8cm);
\foreach \x/\y in {-2.124/1.061,-1.414/1.939,0.5641/2.343,1.901/1.479,2.263/-0.7909,1.386/-1.975,-1.525/-1.863}
{ \shade [ball color=red] (\x,\y) circle (4pt);}
\draw [line width=4pt,->] (2.8,0) -- (4,0);
\node at (-2.124-0.4,1.061) {$\pmb 1$};
\node at (-1.525-0.4,-1.863) {$\pmb n$};

\begin{scope}[xshift=7cm]
\filldraw [cyan!30!white] (0,0) circle (2.4cm);
\draw [ultra thick] (0,0) circle (2.4cm);
\foreach \x/\y in {-2.124/1.061,-1.414/1.939,0.5641/2.343,1.901/1.479,2.263/-0.7909,1.386/-1.975,-1.525/-1.863}
{ \shade [ball color=red] (\x,\y) circle (4pt);}
\node at (-2.124-0.4,1.061) {$\pmb 1$};
\node at (-1.525-0.4,-1.863) {$\pmb n$};
\shade [ball color=red] (0,0) circle (4pt);
\end{scope}

\node at (0,2.4+0.2){};
\node at (0,-2.4-0.2){};

\end{tikzpicture}
} & Cyclohedron $\mathcal{B}_{n-1}$ & $X_0$
\\
\hline
\parbox[c]{0.6\linewidth}{
\begin{tikzpicture}[>=latex,scale=0.5, every node/.style={transform shape}]

\filldraw [cyan!30!white] (0,0) circle (2.4cm);
\filldraw [white] (-1.2,0) circle (0.5cm);
\draw [ultra thick] (0,0) circle (2.4cm);
\draw [ultra thick] (-1.2,0) circle (0.5cm);
\draw [red, ultra thick] (-2,1.3) .. controls (-0.0198, 0.8665) and (-0.0198, -0.8665) .. (-2,-1.3);
\shade [ball color=red] (2.4,0) circle (4pt);
\foreach \x/\y in {0.6755/2.301,2.124/1.117,1.776/-1.627,-0.578/-2.323}
{\shade [ball color=red](\x,\y) circle (4pt);}

\node at (-1.386, 1.953+0.4){$\pmb i+1$};
\node at (-1.386, -1.953-0.4){$\pmb i$};
\shade [ball color=red] (-1.386, 1.953) circle (4pt);
\shade [ball color=red] (-1.386, -1.953) circle (4pt);

\draw [line width=4pt,->] (2.8,0) -- (4,0);

\begin{scope}[xshift=7cm]
\filldraw [cyan!30!white] (0,0) circle (2.4cm);
\draw [ultra thick] (0,0) circle (2.4cm);
\shade [ball color=red] (2.4,0) circle (4pt);
\filldraw [white] (0,0) circle (0.8cm);
\draw [ultra thick] (0,0) circle (0.8cm);
\end{scope}
\begin{scope}[xshift=9.4cm+2.4cm]
\filldraw [cyan!30!white] (0,0) circle (2.4cm);
\draw [ultra thick] (0,0) circle (2.4cm);
\node at (-1.386, 1.953+0.4){$\pmb i+1$};
\node at (-1.386, -1.953-0.4){$\pmb i$};
\shade [ball color=red] (-1.386, 1.953) circle (4pt);
\shade [ball color=red] (-1.386, -1.953) circle (4pt);
\shade [ball color=red] (-2.4,0) circle (4pt);
\foreach \x/\y in {0.6755/2.301,2.124/1.117,1.776/-1.627,-0.578/-2.323,2.4/0}
{\shade [ball color=red](\x,\y) circle (4pt);}
\end{scope}

\end{tikzpicture}

} & Tadpole Factorisation $\mathcal{A}_{n-1} \times \rm{H}_1$ & $X_{(i+1,i+2,\dots,i)}$
\\
\hline
\parbox[c]{0.55\linewidth}{
\begin{tikzpicture}[>=latex,scale=0.5, every node/.style={transform shape}]
\filldraw [cyan!30!white] (0,0) circle (2.4cm);
\filldraw [white] (0,0) circle (0.8cm);
\draw [ultra thick, red] (-1.3,2) -- (-1.3,-2);
\draw [ultra thick] (0,0) circle (2.4cm);
\draw [ultra thick] (0,0) circle (0.8cm);
\foreach \x/\y in {-1.79/1.591,-2.18/1.006,-2.375/0.2954,-2.166/-1.014,-1.706/-1.696,-0.2577/-2.365,1.4/-1.961,2.403/-0.09453,1.957/1.382,-0.2994/2.371}
{
\shade [ball color=red] (\x,\y) circle (4pt);
}
\draw [line width=4pt,->] (2.8,0) -- (4,0);
\node at (-1.789, 1.757+0.3) {$\pmb i$};
\node at (-1.789, -1.757-0.3) {$\pmb j$};
\node at (-0.2981, 2.371+0.3) {$\pmb i+1$};
\node at (-0.2981, -2.371-0.3) {$\pmb j-1$};
\begin{scope}[xshift=7cm]
\filldraw [cyan!30!white] (0,0) circle (2.4cm);
\draw [ultra thick] (0,0) circle (2.4cm);
\shade [ball color=red] (2.4,0) circle (4pt);
\foreach \x/\y in {-1.706/-1.699,-2.18/-1.002,-2.361/0.2941,-2.18/1.005,-1.775/1.59}
{
\shade [ball color=red] (\x,\y) circle (4pt);
}
\node at (-1.789, 1.757+0.3) {$\pmb i$};
\node at (-1.789, -1.757-0.3) {$\pmb j$};
\end{scope}
\begin{scope}[xshift=9.4cm+2.4cm]
\filldraw [cyan!30!white] (0,0) circle (2.4cm);
\draw [ultra thick] (0,0) circle (2.4cm);
\filldraw [white] (0,0) circle (0.8cm);
\draw [ultra thick] (0,0) circle (0.8cm);

\node at (-1.386, 1.953+0.4){$\pmb i+1$};
\node at (-1.386, -1.953-0.4){$\pmb j-1$};
\shade [ball color=red] (-1.386, 1.953) circle (4pt);
\shade [ball color=red] (-1.386, -1.953) circle (4pt);
\shade [ball color=red] (-2.4,0) circle (4pt);
\foreach \x/\y in {0.6755/2.301,2.124/1.117,1.776/-1.627,-0.578/-2.323,2.4/0}
{\shade [ball color=red](\x,\y) circle (4pt);}

\end{scope}

\end{tikzpicture}
} & Factorisation $\mathcal{A}_{m} \times \rm{H}_{n-m}$ & $X_{(j,j+1,\dots,i)}$
\\
\hline
\parbox[c]{0.75\linewidth}{
\begin{tikzpicture}[>=latex,scale=0.5, every node/.style={transform shape}]
\filldraw[cyan!30!white] (0,0) circle (2.4cm);
\filldraw[white] (0,0) circle (0.8cm);
\draw [ultra thick] (0,0) circle (2.4cm);
\draw [ultra thick] (0,0) circle (0.8cm);
\shade [ball color=red] (2.4,0) circle (4pt);
\shade [ball color=red] (-2,1.3) circle (4pt);
\shade [ball color=red] (-2,-1.3) circle (4pt);
\foreach \x/\y in {-1.595/1.786,0.6755/2.301,2.124/1.117,1.776/-1.627,-0.578/-2.323}
{\shade [ball color=red](\x,\y) circle (4pt);}
\draw [red, ultra thick] (-2.4,0) -- (-0.8,0);
\draw [line width=4pt,->] (2.8,0) -- (4,0);
\node at (-2.2-0.3,1.5) {$\pmb i+1$};
\node at (-2.2,-1.5) {$\pmb i$};

\begin{scope}[xshift=7cm]
\filldraw[cyan!30!white] (0,0) circle (2.4cm);
\filldraw[white] (-1.6,0) circle (0.8cm);
\draw [ultra thick] (-1.6,0) circle (0.8cm);
\draw [ultra thick] (0,0) circle (2.4cm);
\shade [ball color=red] (2.4,0) circle (4pt);
\shade [ball color=red] (-2,1.3) circle (4pt);
\shade [ball color=red] (-2,-1.3) circle (4pt);
\draw [line width=4pt,->] (2.8,0) -- (4,0);
\foreach \x/\y in {-1.595/1.786,0.6755/2.301,2.124/1.117,1.776/-1.627,-0.578/-2.323}
{\shade [ball color=red](\x,\y) circle (4pt);}
\node at (-2.2-0.3,1.5) {$\pmb i+1$};
\node at (-2.2,-1.5) {$\pmb i$};

\end{scope}
\begin{scope}[xshift=14cm]
\filldraw [cyan!30!white] (0,0) circle (2.4cm);
\draw [ultra thick] (0,0) circle (2.4cm);
\shade [ball color=red] (2.4,0) circle (4pt);
\shade [ball color=red] (-2,1.3) circle (4pt);
\shade [ball color=red] (-2,-1.3) circle (4pt);
\shade [ball color=red] (1.386, 1.953) circle (4pt);
\shade [ball color=red] (1.386, -1.953) circle (4pt);
\node at (-1.985-0.4, 1.298){$\pmb +$};
\node at (-1.985-0.4, -1.298){$\pmb -$};
\node at (1.386+0.2, 1.953+0.4){$\pmb i+1$};
\node at (1.386, -1.953-0.4){$\pmb i$};
\foreach \x/\y in {2.11/1.159,2.333/0.6297,2.389/-0.3592,2.221/-0.9441}{\shade [ball color=red](\x,\y) circle (4pt);}
\end{scope}
\end{tikzpicture}

} & Cut Associahedron $\mathcal{A}_{n-1}$ & $X_i$
\\
\hline
\end{tabular}
}
\caption{List of the arcs and corresponding facets of the Halohedron, which are also labelled by dual planar variables.}
\label{tab:facethaloedron}
\end{table}
\FloatBarrier

\begin{figure}[!htb]
\centering
\minipage{0.33\textwidth}
\centering
  \begin{tikzpicture}[scale=3.5]
    \def\unit{.25 cm};
    \def\smallunit{.05 cm};
    \draw [ultra thick, gray] (-\unit,0) -- (\unit,0);
    \draw [ultra thick, gray] (-\unit,0) -- (-2*\unit,-\unit);
    \draw [ultra thick, gray] (-\unit,0) -- (-\unit,2*\unit);
    \filldraw[draw=black, dashed, fill=black, opacity=0.8, ultra thick] (0, \unit) ++ (-3*\smallunit,0) -- ++ (5 *\smallunit,2*\smallunit) -- ++ (-2*\smallunit,-5*\smallunit) -- cycle;
    \shade[left color = green!60!black, opacity = 0.6,draw = black, ultra thick] (-2*\unit,\unit) -- (0, \unit) -- (\unit, 2* \unit) -- (-\unit, 2* \unit) -- cycle;
    \shade[left color = green!60!black, opacity = 0.6,draw = black, ultra thick](-2*\unit,\unit) -- (0, \unit) -- (0, -\unit) -- (-2*\unit, -\unit)--cycle;
    \shade[left color = green!60!black, opacity = 0.6,draw = black, ultra thick] (0, \unit) -- (0, -\unit) -- (\unit,0)-- (\unit,2*\unit) -- cycle;
    \draw[ultra thick, ->] (\unit*1.5,\unit/2) -- ++ (\unit,0);

  \end{tikzpicture}
\endminipage\hfill
\minipage{0.33\textwidth}
\centering
  \begin{tikzpicture}[scale=3.5]
    \def\unit{.25 cm};
    \def\smallunit{.05 cm};
    \def\A{(-2*\unit,\unit)++ (.75*\smallunit,.75*\smallunit)};
    \def\B{(\unit,2*\unit) ++ (-.75*\smallunit,0)};
    \def\C{(0,-\unit) ++ (-\smallunit,0)};
    \draw [ultra thick, gray] (-\unit,0) -- (\unit,0);
    \draw [ultra thick, gray] (-\unit,0) -- (-2*\unit,-\unit);
    \draw [ultra thick, gray] (-\unit,0) -- (-\unit,2*\unit);
    \filldraw[draw=black, dashed, fill=black, opacity=0.8, ultra thick] \A -- ++ (\unit+3*\smallunit,0) -- ++ (-0.75*\smallunit,-1.75*\smallunit) -- ++ (-\unit-3*\smallunit,0) -- cycle;
    \filldraw[draw=black, dashed, fill=black, opacity=0.8, ultra thick] \B -- ++ (-3.75*\smallunit,-3.75*\smallunit) -- ++ (0.9*\smallunit,-0.9*\smallunit) -- ++ (3.75*\smallunit,3.75*\smallunit) -- cycle;
    \filldraw[draw=black, dashed, fill=black, opacity=0.8, ultra thick] \C -- ++ (0,8*\smallunit) -- ++ (2*\smallunit,\smallunit) -- ++ (0,-8*\smallunit) -- cycle;
        \shade[left color = green!60!black, opacity = 0.6,draw = black, ultra thick](0, \unit) ++ (-3*\smallunit,0) -- ++ (5 *\smallunit,2*\smallunit) -- ++ (-2*\smallunit,-5*\smallunit) -- cycle;
    \shade[left color = green!60!black, opacity = 0.6,draw = black, ultra thick] (0, \unit) ++ (-3*\smallunit,0) -- ++ (5 *\smallunit,2*\smallunit)-- (\unit, 2* \unit) -- (-\unit, 2* \unit) -- (-2*\unit,\unit);
    \shade[left color = green!60!black, opacity = 0.6,draw = black, ultra thick] (0, \unit) ++ (-3*\smallunit,0) -- ++ (3*\smallunit,-3*\smallunit)-- (0, -\unit) -- (-2*\unit, -\unit)-- (-2*\unit,\unit) -- cycle;
    \shade[left color = green!60!black, opacity = 0.6,draw = black, ultra thick]
    (0, \unit) ++ (-3*\smallunit,0) ++ (5 *\smallunit,2*\smallunit) -- (\unit, 2*\unit) --  (\unit,0) -- (0,-\unit) -- 
    (0, \unit-3*\smallunit) -- cycle;
    \draw[ultra thick, ->] (\unit*1.5,\unit/2) -- ++ (\unit,0);
  \end{tikzpicture}
\endminipage\hfill
\minipage{0.33\textwidth}
\centering
  \begin{tikzpicture}[scale=0.35,every node/.style={transform shape}]
    \shade [left color=green!40!white, right color=green!10!white] (0,10) -- (3.464105, 8) -- (4.33013, 6.5) -- (4.33013, 2.5) -- (0.866025, 0.5) -- (-0.866025, 0.5) -- (-4.33013, 2.5) -- (-4.33013, 6.5) -- (-3.464105, 8) -- cycle;
    \shade [left color=green!60!white, right color=green!40!white] (-0.866025, 6.5) -- (-3.464105, 8) -- (-4.33013, 6.5) -- (-1.73205, 5) -- cycle;
    \shade [left color=green!40!white, right color=green!20!white] (0.866025, 6.5) -- (3.464105, 8) -- (4.33013, 6.5) -- (1.73205, 5) -- cycle;
    \shade [left color=green!50!white, right color=green!30!white] (0.866025, 0.5) -- (-0.866025, 0.5) -- (-0.866025, 3.5) -- (0.866025, 3.5) -- cycle;
    \shade [left color=green!70!white, right color=green!40!white] (0.866025, 6.5) -- (-0.866025, 6.5) -- (-1.73205, 5) -- (-0.866025, 3.5) -- (0.866025, 3.5) -- (1.73205, 5) -- cycle;
    \draw [ultra thick] (0,10) -- (3.464105, 8) -- (4.33013, 6.5) -- (4.33013, 2.5) -- (0.866025, 0.5) -- (-0.866025, 0.5) -- (-4.33013, 2.5) -- (-4.33013, 6.5) -- (-3.464105, 8) -- cycle;
    \draw [ultra thick] (-0.866025, 6.5) -- (-3.464105, 8) -- (-4.33013, 6.5) -- (-1.73205, 5) -- cycle;
    \draw [ultra thick] (0.866025, 6.5) -- (3.464105, 8) -- (4.33013, 6.5) -- (1.73205, 5) -- cycle;
    \draw [ultra thick] (0.866025, 0.5) -- (-0.866025, 0.5) -- (-0.866025, 3.5) -- (0.866025, 3.5) -- cycle;
    \draw [ultra thick] (0.866025, 6.5) -- (-0.866025, 6.5) -- (-1.73205, 5) -- (-0.866025, 3.5) -- (0.866025, 3.5) -- (1.73205, 5) -- cycle;
    \draw [dashed,opacity=0.5] (-4.33013, 2.5) -- (0,5); 
    \draw [dashed,opacity=0.5] (4.33013, 2.5) -- (0,5); 
    \draw [dashed,opacity=0.5] (0, 10) -- (0,5); 
    \shade [ball color=red] (0,10) circle (3pt);
    \shade [ball color=red]  (3.464105, 8) circle (3pt);
    \shade [ball color=red] (4.33013, 6.5) circle (3pt);
    \shade [ball color=red] (4.33013, 2.5) circle (3pt);
    \shade [ball color=red] (0.866025, 0.5) circle (3pt);
    \shade [ball color=red] (-0.866025, 0.5) circle (3pt);
    \shade [ball color=red] (-4.33013, 2.5) circle (3pt);
    \shade [ball color=red] (-4.33013, 6.5) circle (3pt);
    \shade [ball color=red] (-3.464105, 8) circle (3pt);
    \shade [ball color=red, opacity=0.5] (0, 5) circle (3pt);
    \shade [ball color=red] (0.866025, 6.5) circle (3pt);
    \shade [ball color=red] (-0.866025, 6.5) circle (3pt);
    \shade [ball color=red] (-1.73205, 5) circle (3pt);
    \shade [ball color=red] (1.73205, 5) circle (3pt);
    \shade [ball color=red] (-0.866025, 3.5) circle (3pt);
    \shade [ball color=red] (0.866025, 3.5) circle (3pt);
    \filldraw [cyan!30!white] (7,8) circle (1.2cm);
    \filldraw [white] (6.5,8) circle (0.4cm);
    \draw [ultra thick, red] (5.9,8.5) .. controls (7.5,8.9) and (7.5,7.1) .. (5.9,7.5);
    \draw [ultra thick] (7,8) circle (1.2cm);
    \draw [ultra thick] (6.5,8) circle (0.4cm);
    \filldraw [red] (7,9.2) circle (2pt);
    \node at (7,9.5) {\pmb {3}};
    \filldraw [red] (7,6.8) circle (2pt);
    \node at (7,6.5) {\pmb {1}};
    \filldraw [red] (8.2,8) circle (2pt);
    \node at (8.5,8) {\pmb {2}};
    \filldraw [cyan!30!white] (8.5,5) circle (1.2cm);
    \filldraw [white] (8,5) circle (0.4cm);
    \draw [ultra thick, red] (8.9,6.1) -- (8.9,3.9);
    \draw [ultra thick] (8.5,5) circle (1.2cm);
    \draw [ultra thick] (8,5) circle (0.4cm);
    \filldraw [red] (7.3,5) circle (2pt);
    \node at (7.1,5) {\pmb {1}};
    \filldraw [red] (9.3,5.9) circle (2pt);
    \node at (9.3,6.2) {\pmb {3}};
    \filldraw [red] (9.3,4.1) circle (2pt);
    \node at (9.3,3.8) {\pmb {2}};
    \filldraw [cyan!30!white] (6.6,2) circle (1.2cm);
    \filldraw [white] (6.6,2) circle (0.4cm);
    \draw [ultra thick, red] (6.6,1.6) -- (6.6,0.8);
    \draw [ultra thick] (6.6,2) circle (1.2cm);
    \draw [ultra thick] (6.6,2) circle (0.4cm);
    \filldraw [red] (6.6,3.2) circle (2pt);
    \node at (6.6,3.5) {\pmb {3}};
    \filldraw [red] (5.4,2) circle (2pt);
    \node at (5.2,2) {\pmb {1}};
    \filldraw [red] (7.8,2) circle (2pt);
    \node at (8,2) {\pmb {2}};
    \node (a) at (7.4,4.7) {};
    \node (b) at (3.5,4) {};
    \draw[blue] [-latex, bend left, ultra thick] (a) to (b);
    \node (c) at (5.8,8.2) {};
    \node (d) at (3.5,7) {};
    \draw[blue] [-latex, bend right, ultra thick] (c) to (d);
    \node (e) at (5.4,1.8) {};
    \node (f) at (2.5,1.5) {};
    \draw[blue] [-latex, bend left, ultra thick] (e) to (f);
    \node (g) at (-3.5,9.5){};
    \node (h) at (0,5.8) {};
    \draw[blue] [-latex, bend left, ultra thick] (g) to (h);
    \filldraw [cyan!30!white] (-5,9.5) circle (1.2cm);
    \filldraw [white] (-5,9.5) circle (0.4cm);
    \draw [ultra thick] (-5,9.5) circle (1.2cm);
    \draw [ultra thick, red] (-5,9.5) circle (0.4cm);
    \filldraw [red] (-5,10.7) circle (2pt);
    \filldraw [red] (-5-1.2,9.5) circle (2pt);
    \filldraw [red] (-5+1.2,9.5) circle (2pt);
    \node (i) at (-5+0.2,10.7+0.2) {\pmb {1}};
    \node (n) at (-5-1.2-0.2,9.5+0.2){\pmb {2}};
    \node (m) at (-5+1.2+0.2,9.5+0.2){\pmb {3}};   
  \end{tikzpicture}
\endminipage\hfill
\caption{The sequence of truncations that produces the 3-dimensional Halohedron.}
\label{fig:halohedrontruncations}
\end{figure}
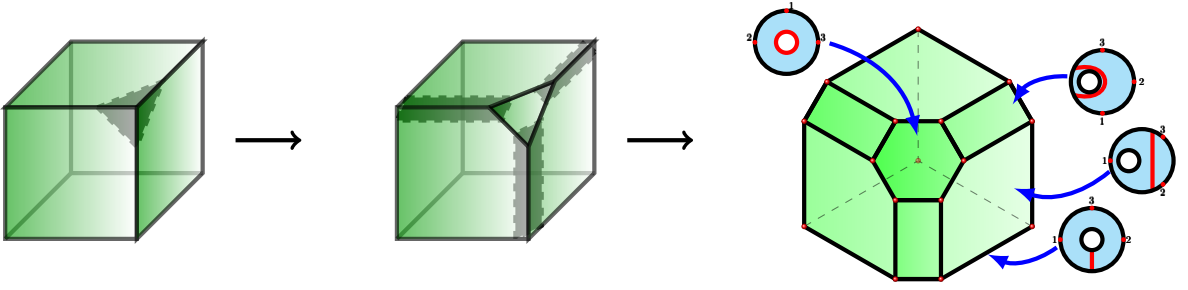

In order to find the correct form of the functions $X_I$, we implement the convex realisation of the Halohedron described in \cite{devadoss}, which is obtained by iterated truncations of an $n$-dimensional cube. We center one of the corners of the cube at the origin of the space $X$, so that the coordinates $X_i$ become the face variables of $n$ of the facets of the cube. Next, we introduce functions $X_{(i,i+1)} = \epsilon_{i} - X_i$, for some positive constant $\epsilon_i$. The positive regions $X_{i} \ge 0$ and $X_{(i,i+1)} \ge 0$ define the initial cube, which is to be truncated at the intersection of the faces $X_{(i,i+1)}$ in order of increasing dimensions. The first truncation happens at the vertex where all the facets $X_{(i,i+1)}$ meet and it is implemented by considering the function
\begin{align}
X_0 = \sum_{a=1}^n X_{(a,a+1)} - \epsilon_0,
\label{eq:x0}
\end{align}
where $\epsilon_0$ is a new positive constant. Requiring $X_0 \ge 0$ shaves off the vertex where all the $X_{i,i+1}$ are zero, creating a new facet - which is a $(n-1)$-simplex - at the end of all the truncations such facet will be cutted into the cyclohedral facet of the Halohedron. Similarly one truncates all the one dimensional faces given by the intersection of the faces $X_{a,a+1}$ for $a \in (i, i+1,\dots, i-1)$, by introducing functions
\begin{align}
X_{(i,i+1,\dots,i-1)} := \sum_{a=i}^{i-1} X_{(a,a+1)} - \epsilon_{(i,i+1,\dots,i-1)},
\end{align}
and demanding them to be positive. The truncations easily generalize to every dimension, for every subset $I \subset (1,2,\dots,n)$ of cyclically consecutive indices and of cardinality $|I| \ge 3$ we consider a function
\begin{align}
X_I = \sum_{a \in I'} X_{(a,a+1)} - \epsilon_I,
\label{eq:halovar}
\end{align}
where $I'$ is obtained from $I$ by dropping the last element and $\epsilon_I$ is a positive constant. The variables $X_i$, $X_{i,i+1}$,$X_0$ and $X_{I}$ together span the whole set of facets of the Halohedron $\rm{H}_n$, and the region where they are simultaneously positive gives a convex realisation of it, an example is shown in Fig. \ref{fig:halohedrontruncations}. Finally, we remark the constants $\epsilon_I$ cannot be chosen arbitrarily. The reason is that they modulate the depth of the truncations which must not be too deep, for example $\epsilon_0$ must be smaller than $\epsilon_{i}$ or the facet $X_0$ created by the truncation will touch the facet $X_i$.

\section{The Canonical form and the Integrand}
\label{The Canonical form and the Integrand}

We now study the canonical form of the Halohedron we introduced in the previous section.

In general, when working with a $n$-dimensional simple polytope $\mathcal{P}$ (i.e. one whose vertices are adjacent to exactly $n$ facets) the canonical form can be written as
\begin{align}
\Omega_{\mathcal{P}} = \sum_{v \in \mathrm{vertices}} \mathrm{sign}(v) \bigwedge\limits_{f\ |\ v \in f} d\mathrm{log}(X_f),
\label{eq:formsimple}
\end{align}
the sum is taken over all vertices $v$ of $\mathcal{P}$, the wedge product over all facets $f$ meeting in $v$, each of which is defined and labelled by a linear equation $\{X_f = 0\}$.
The ordering of the facets $X_f$ and the relative signs $\mathrm{sign}(v)$ must be chosen so that the form is \emph{projective}, i.e. invariant under $X_f \to \alpha(X) X_f$. In practice this is guaranteed by the following \emph{mutation rule}. Suppose $v$ and $v'$ are adjacent on the same 1-dimensional boundary $E$ of the polytope. Then they are given by the intersection of two sets of facets which have all but two elements equal. Let us call $X_f$ and $X_{f'}$ these two elements. Once the facets in the wedge product are ordered so that $X_f$ and $X_{f'}$ are in the same position, we must have $\mathrm{sign}(v) = - \mathrm{sign}(v')$. This has to be, because we can take iterate residues of $\Omega_{\mathcal{P}}$ until we are left with $\Omega_{E} = d\mathrm{log}(X_f) \pm d\mathrm{log}(X_{f'})$, and so we see that we need a ``-'' in order to avoid a double pole at infinity.

\begin{figure}[!htbp]

\begin{center}

\begin{tikzpicture}[>=latex,scale=4]

\draw [<->] (-.1,.5) -- (.1,.5);
\draw [<->] (-.1,-.5) -- (.1,-.5);

\begin{scope}[xshift=-.5cm,yshift=+.5cm]
\begin{scope}[scale=0.25]

\draw[thick] (-.5,0) -- (.5,0);

\draw[thick] (-1,1) -- (-.5,0) -- (-1,-1);
\draw[thick] (-1,1) -- (-.5,0) -- (-1,-1);
\draw[thick] (1,1) -- (.5,0) -- (1,-1);
\draw[thick] (1,1) -- (.5,0) -- (1,-1);

\foreach \t in {0,90,180,270}
{
\begin{scope}[rotate = \t]
\draw[red,dashed] (1,1) -- + (.4,-.4) --  +(-.4,.4);
\end{scope}
};

\end{scope}
\end{scope}

\begin{scope}[xshift=.5cm,yshift=+.5cm]
\begin{scope}[xscale=0.25,yscale=0.25,rotate=90]

\draw[thick] (-.5,0) -- (.5,0);

\draw[thick] (-1,1) -- (-.5,0) -- (-1,-1);
\draw[thick] (-1,1) -- (-.5,0) -- (-1,-1);
\draw[thick] (1,1) -- (.5,0) -- (1,-1);
\draw[thick] (1,1) -- (.5,0) -- (1,-1);

\foreach \t in {0,90,180,270}
{
\begin{scope}[rotate = \t]
\draw[red,dashed] (1,1) -- + (.4,-.4) --  +(-.4,.4);
\end{scope}
};

\end{scope}
\end{scope}

\begin{scope}[xshift=-.5cm,yshift=-.5cm]
\begin{scope}[xscale=0.25,yscale=0.25]

\filldraw[cyan!30!white] (0,0) circle (1cm);
\draw[thick,black] (0,0) circle (1cm);
\draw[thick,red,dashed] (-1,0) -- (0,-1) -- (1,0) -- (0,1) -- (-1,0);
\draw[thick,black] (0,-1) -- (0,1);

\end{scope}
\end{scope}

\begin{scope}[xshift=.5cm,yshift=-.5cm]
\begin{scope}[xscale=0.25,yscale=0.25,rotate=90]

\filldraw[cyan!30!white] (0,0) circle (1cm);
\draw[thick,black] (0,0) circle (1cm);
\draw[thick,red,dashed] (-1,0) -- (0,-1) -- (1,0) -- (0,1) -- (-1,0);
\draw[thick,black] (0,-1) -- (0,1);

\end{scope}
\end{scope}

\end{tikzpicture}

\caption{The usual mutation swaps an s-channel for a t-channel in a subdiagram. The move is represented also in terms of arcs on the annulus}
\label{fig:mutazionesolita}

\end{center}
\end{figure}
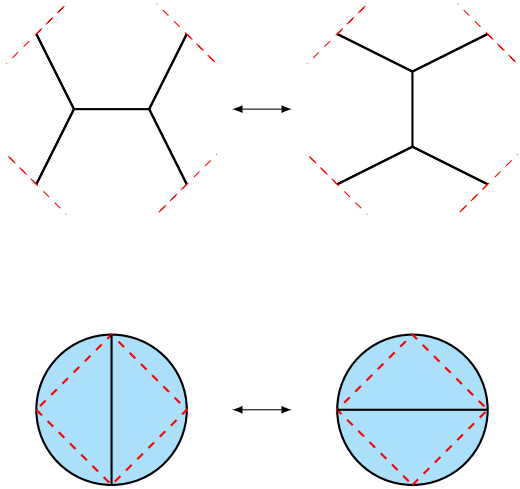

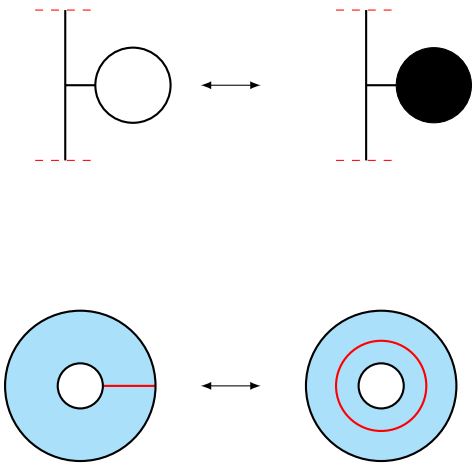
\begin{figure}[!htbp]
\begin{center}

\begin{tikzpicture}[>=latex,scale=4]

\draw [<->] (-.1,.5) -- (.1,.5);
\draw [<->] (-.1,-.5) -- (.1,-.5);

\begin{scope}[xshift=-.5cm,yshift=+.5cm]
\begin{scope}[scale=0.25]

\draw[red,dashed] (-.4-.2,1) -- (.4-.2,1);
\draw[red,dashed] (-.4-.2,-1) -- (.4-.2,-1);

\draw[thick] (-.2,-1) -- (-.2,1);
\draw[thick] (-.2,0) -- (.2,0);
\draw[thick] (.7,0) circle (.5);

\end{scope}
\end{scope}

\begin{scope}[xshift=.5cm,yshift=+.5cm]
\begin{scope}[xscale=0.25,yscale=0.25]

\draw[red,dashed] (-.4-.2,1) -- (.4-.2,1);
\draw[red,dashed] (-.4-.2,-1) -- (.4-.2,-1);

\draw[thick] (-.2,-1) -- (-.2,1);
\draw[thick] (-.2,0) -- (.2,0);
\filldraw[black] (.7,0) circle (.5);

\end{scope}
\end{scope}

\begin{scope}[xshift=-.5cm,yshift=-.5cm]
\begin{scope}[xscale=0.25,yscale=0.25]

\filldraw[cyan!30!white] (0,0) circle (1cm);
\draw[red,thick] (.3,0) -- (1,0);
\draw[thick] (0,0) circle (1cm);
\filldraw[white] (0,0) circle (0.3cm);
\draw[thick] (0,0) circle (0.3cm);

\end{scope}
\end{scope}

\begin{scope}[xshift=.5cm,yshift=-.5cm]
\begin{scope}[xscale=0.25,yscale=0.25,rotate=90]

\filldraw[cyan!30!white] (0,0) circle (1cm);
\draw[red,thick] (0,0) circle (.6cm);
\draw[thick] (0,0) circle (1cm);
\filldraw[white] (0,0) circle (0.3cm);
\draw[thick] (0,0) circle (0.3cm);

\end{scope}
\end{scope}

\end{tikzpicture}

\caption{The new mutation swaps IR and UV tadpoles, the latter are distinguished by coloring the tadpole in black. In terms of arcs on the annulus, the mutation swaps a cut arc with the UV arc.}
\label{fig:nuovamutazione}

\end{center}
\end{figure}

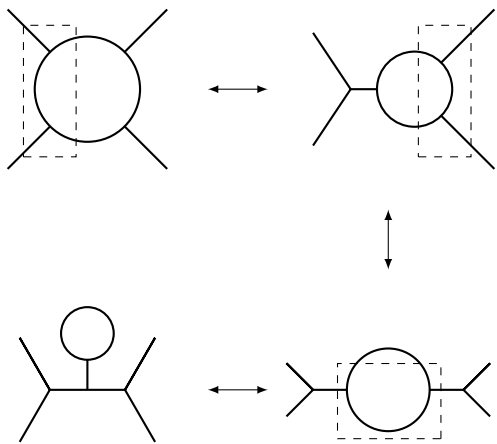
\begin{figure}[!htbp]

\begin{center}

\begin{tikzpicture}[>=latex,scale=4]


\draw [<->] (-.1,.5) -- (.1,.5);
\draw [<->] (-.1,-.5) -- (.1,-.5);
\draw [<->] (.5,.1) -- (.5,-.1);

\begin{scope}[xshift=-.5cm,yshift=+.5cm]
\begin{scope}[scale=0.25]

\draw[thick] (0,0) circle (.7cm);

\foreach \t in {0,90,180,270}{\draw[thick] (45 + \t:.7) -- (45+\t:1.5);}

 \begin{scope}[xshift=-.5cm,yshift=+.5cm]
 \begin{scope}[scale=0.35]

   \draw[dashed] (-1,-4) -- (1,-4) -- (1,1) -- (-1,1) -- (-1,-4);

 \end{scope}
 \end{scope}

\end{scope}
\end{scope}

\begin{scope}[xshift=.5cm,yshift=+.5cm]
\begin{scope}[xscale=0.25,yscale=0.25]

\draw[thick] (-1,.75) -- (-.5,0) -- (-1,-.75);
\draw[thick] (.35,0) + (180:.5) -- (-.5,0);
\draw[thick] (.35,0) circle (.5cm);
\draw[thick] (.35,0) +(45:.5) -- +(45:1.5);
\draw[thick] (.35,0) +(-45:.5) -- +(-45:1.5);

 \begin{scope}[xshift=.75cm,yshift=.5cm]
 \begin{scope}[scale=0.35]

   \draw[dashed] (-1,-4) -- (1,-4) -- (1,1) -- (-1,1) -- (-1,-4);

 \end{scope}
 \end{scope}

\end{scope}
\end{scope}

\begin{scope}[xshift=-.5cm,yshift=-.5cm]
\begin{scope}[xscale=0.25,yscale=0.25]

\draw[thick](-.5,0) -- (.5,0);

\draw[thick] (.5,0) -- ++ (60:.8) -- ++ (60:-.8) -- ++ (-60:.8);
\draw[thick] (-.5,0) -- ++ (120:.8) -- ++ (120:-.8) -- ++ (-120:.8);

\draw[thick] (0,.4) -- (0,0);
\draw[thick] (0,.75) circle (.35cm);

\end{scope}
\end{scope}

\begin{scope}[xshift=.5cm,yshift=-.5cm]
\begin{scope}[xscale=0.25,yscale=0.25,rotate=0]

\draw[thick] (0,0) circle (0.55cm);
\draw[thick] (0,0) + (180:0.55) -- (-1,0);
\draw[thick] (0,0) + (0:0.55) --  (1,0);

\draw[thick] (1,0) -- ++ (45:.5) -- ++ (45:-.5) -- ++ (-45:.5);
\draw[thick] (-1,0) -- ++ (135:.5) -- ++ (135:-.5) -- ++ (-135:.5);

 \begin{scope}[xshift=-.4cm,yshift=-.15cm]
 \begin{scope}[yscale=2]
 \begin{scope}[xscale=1.1]
 \begin{scope}[scale=0.25,rotate=90]

   \draw[dashed] (-1,-4) -- (1,-4) -- (1,1) -- (-1,1) -- (-1,-4);

 \end{scope}
 \end{scope}
 \end{scope}
 \end{scope}

\end{scope}
\end{scope}

\end{tikzpicture}

\caption{A succession of mutations on planar loop diagrams}
\label{fig:mutationsequence}

\end{center}
\end{figure}

We can label the vertices of both the Associahedron and the Halohedron in terms of Feynman diagrams. Then, two vertices happen to be adjacent to the same edge $E$ if and only if their Feynman diagrams are related by an $s/t$-channel swap (or \emph{mutation}) as in Fig. \ref{fig:mutazionesolita}.
For the Halohedron, the mutation rule is generalised to include a swap of an IR tadpole with a UV one\footnote{We recall that, as described in \cite{Salvatori:2018fjp}, the labelling of the vertices of the Halohedron in terms of Feynman diagrams requires to introduce two kind of tadpoles which were dubbed IR/UV.}, see Fig. \ref{fig:nuovamutazione}.
Example of mutations for 1-loop diagrams are shown in Fig. \ref{fig:mutationsequence}, note how bubbles turn into tadpoles, and factorisation channels emerge from loop propagators.
Therefore, in the case of the Halohedron \eqref{eq:formsimple} becomes
\begin{align}
\label{eq:canonical}
\Omega_{\mathrm{H}_n} = \sum_{g} \mathrm{sign}(g) \bigwedge_{I \in g} d\mathrm{log}(X_I),
\end{align}
$I$ runs over all the propagators of a diagram $g$, and $\mathrm{sign}(g)$ is fixed by mutating every diagram from a chosen reference one.
Crucially, since our functions $X$ are not constrained by the usual momentum-conservation relations, \emph{every} diagram contributes to $\Omega_{\mathrm{H}_n}$: IR/UV tadpoles, internal and external bubbles as well.
We would like to express all the forms $d\mu_g = \bigwedge\limits_{I \in g} dX_I$ appearing in \eqref{eq:canonical} in terms of a single one, so that we can extract from $\Omega_{\mathrm{H}_n}$ the rational function $\underline{\Omega}_{\rm{H}_n}$ which we will interpret as the amplitude.

In the Associahedron case, this was done by using physical propagators in lieu of $X$. Because of momentum conservation, they have to satisfy the \emph{7-term identity}
$$ s_{I_1} + s_{I_2} + s_{I_3} + s_{I_4} = s_{I_1 \cup I_2} + s_{I_1 \cup I_3} + s_{I_1 \cup I_4}.$$
which holds for any partition of the set $\{1, \dots, n\}$ into four sets $I_{1,2,3,4}$ of adjacent indices.
In fact, this identity is \emph{equivalent} to momentum conservation: if the propagators are thought of as abstract variables, the 7-term identity is sufficient to restrict them to the physical sub-space where all relations among propagators hold\cite{Arkani-Hamed:2017mur}.
The 7-term identity, together with the constraints $k_{i,j} = c_{i,j}$ where $c_{i,j}$ is a positive constant and $(i,j)$ non adjacent indices, implies
$$ds_{I_1 \cup I_2} = - ds_{I_1 \cup I_4} + ds_{I_1} + ds_{I_2} + ds_{I_3} + ds_{I_4}.$$
Since for mutated diagrams the terms $ds_{I_i}$ are shared, we are allowed to exchange an $s$-channel for a $t$-channel in the measure of diagrams. We pick a relative sign which is balanced by the one implied by the mutation rule, and thus we can express the canonical form of the Associahedron as
\begin{align*}
\Omega_{\mathcal{A}_{n-3}} = d\mu_{g^*} \sum_g \frac{1}{\prod\limits_{I \in g} X_I},
\end{align*}
where $g^*$ is the reference graph chosen. It is clear that the rational function $\underline{\Omega}_{\mathcal{A}_{n-3}}$, obtained by stripping of the measure $d\mu_{g^*}$ from $\Omega_{\mathcal{A}_{n-3}}$, is a tree level amplitude.

We would like this story to repeat for the Halohedron, but if we imposed the $7$-term identity we would end up again with physical propagators, so that the contribution of internal bubbles would disappear from \eqref{eq:canonical}. In fact, we do not need to impose any constraints on the variables $X$. The functions $X$ defined in \eqref{eq:halovar} are such that 
\begin{align}
\label{eq:mutationdx}
dX \wedge \dots = - dX' \wedge \dots
\end{align}
where $X$ and $X'$ denotes the distinct propagators of two mutated diagrams, whose shared propagators are represented by the dots. We now prove this statement studying case by case the various types of mutations.

\noindent
\emph{Cut/Factorisation adjacency}. Consider two diagrams such as in figure.

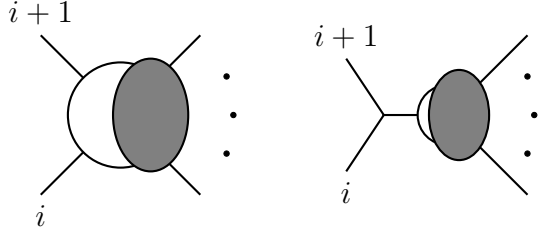
\begin{figure}[h!]
\begin{centering}

\begin{tikzpicture}[>=latex,scale=4][every node/.style={transform shape}]

\begin{scope}[xshift=-.5cm,yshift=+.5cm]
\begin{scope}[scale=0.25]

\draw[thick] (0,0) circle (.7cm);
\filldraw[black] (20:1.5) circle (1pt);
\filldraw[black] (0:1.5) circle (1pt);
\filldraw[black] (-20:1.5) circle (1pt);

\foreach \t in {0,270}{\draw[thick] (45 + \t:.7) -- (45+\t:1.5);}
\draw[thick] (135:.7) -- (135:1.5) node[pos=1,above]{\large $i+1$};
\draw[thick] (-135:.7) -- (-135:1.5) node[pos=1,below]{\large $i$};

\draw[thick,fill=gray] (.4,0) ellipse (.5cm and .75cm);

\end{scope}
\end{scope}

\begin{scope}[xshift=.5cm,yshift=+.5cm]
\begin{scope}[xscale=0.25,yscale=0.25]

\draw[thick] (-1,.75) -- (-.5,0) node(a)[pos=0,above]{\large $i+1$} -- (-1,-.75) node(a)[pos=1,below]{\large $i$};
\draw[thick] (.35,0) + (180:.4) -- (-.5,0);
\draw[thick] (.35,0) circle (.4cm);
\draw[thick] (.35,0) +(45:.5) -- +(45:1.5);
\draw[thick] (.35,0) +(-45:.5) -- +(-45:1.5);

\draw[thick,fill=gray] (.5,0) ellipse (.4cm and .6cm);

\filldraw[black] (20:1.5) circle (1pt);
\filldraw[black] (0:1.5) circle (1pt);
\filldraw[black] (-20:1.5) circle (1pt);

\end{scope}
\end{scope}
\end{tikzpicture}
\label{cutfactorisation}

\end{centering}
\end{figure}
\noindent
The corresponding measures are
\begin{align*} 
dX_i \wedge \dots \quad \mathrm{and} \quad dX_{(i,i+1)} \wedge \dots,
\end{align*}
for the diagram on the left and on the right, respectively. Since we have $X_{(i,i+1)} = \epsilon_i - X_i$, \eqref{eq:mutationdx} trivially holds.

\newpage
\noindent
\emph{IR/UV adjacency}. The situation is slightly more complicate for a pair of IR and UV tadpoles.
\noindent
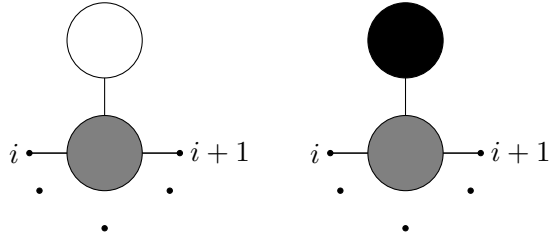
\begin{figure}[h!]
\begin{centering}

\begin{tikzpicture}[>=latex,scale=4][every node/.style={transform shape}]

\begin{scope}[xshift=-.5cm,yshift=+.5cm]
\begin{scope}[scale=0.25]

\foreach \t in {0,-30,-90,210,180}{
\filldraw[black] (-.5,0)+(\t:1) circle (1pt);
}
\foreach \t in {0,180,90}{
\draw[black] (-.5,0)+(\t:1) -- (-.5,0);
}
\draw[black] (-.5,0)+(180:1) -- (-.5,0) node (a)[pos=0,left]{$i$};
\draw[black] (-.5,0)+(0:1) -- (-.5,0) node (a)[pos=0,right]{$i+1$};

\draw(-.5,0)+(90:1.5) circle (.5cm);
\draw[fill=gray] (-.5,0) circle (0.5cm);

\end{scope}
\end{scope}

\begin{scope}[xshift=.5cm,yshift=+.5cm]
\begin{scope}[xscale=0.25,yscale=0.25]

\foreach \t in {0,-30,-90,210,180}{
\filldraw[black] (-.5,0)+(\t:1) circle (1pt);
}
\foreach \t in {0,180,90}{
\draw[black] (-.5,0)+(\t:1) -- (-.5,0);
}
\draw[black] (-.5,0)+(180:1) -- (-.5,0) node (a)[pos=0,left]{$i$};
\draw[black] (-.5,0)+(0:1) -- (-.5,0) node (a)[pos=0,right]{$i+1$};

\draw[fill=black](-.5,0)+(90:1.5) circle (.5cm);
\draw[fill=gray] (-.5,0) circle (0.5cm);

\end{scope}
\end{scope}
\end{tikzpicture}

\label{IRUV}

\end{centering}
\end{figure}
\FloatBarrier
\noindent
 The two diagrams contribution to the canonical form are
\begin{align*}
&d X_i \wedge dX_{(i+1,i+2,\dots,i)} \wedge \dots \\ 
&d X_{0} \wedge dX_{(i+1,i+2,\dots,i)} \wedge \dots,
\end{align*}
this time we highlighted one of the shared propagators, which carries the variable $$dX_{(i+1,i+2,\dots,i)} = - d(X_{i+1} + dX_{i+2} + \dots + dX_{i-1}).$$
Keeping in mind that we are under a wedge with this factor, we can write
\begin{align*}
dX_i \wedge dX_{(i+1,i+2,\dots,i)} &= d(X_i + dX_{i+1} + \dots + dX_{i-1}) \wedge dX_{(i+1,i+2,\dots,i)} \\
&= - dX_0 \wedge dX_{(i+1,i+2,\dots,i)}
\end{align*}
\noindent
\emph{Tadpole/Bubble adjacency}. This adjacency involves two diagrams as in figure.

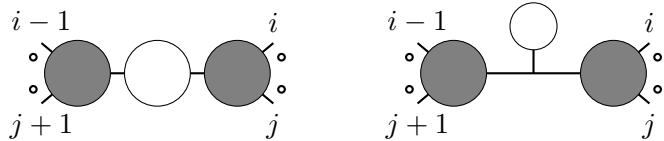
\begin{figure}[h!]
\begin{centering}

\begin{tikzpicture}[>=latex,scale=5][every node/.style={transform shape}]

\begin{scope}[xshift=-.5cm,yshift=+.5cm]
\begin{scope}[scale=0.25]

\foreach \t in {-20,20}{
\draw[thick](.85,0) +(\t:.5) circle (1pt) ;
}
\foreach \t in {180-20,180+20}{
\draw[thick](-.85,0) +(\t:.5) circle (1pt) ;
}

\draw[thick](.85,0) -- +(40:.5) node [pos=1,above]{$i$};
\draw[thick](.85,0) -- +(-40:.5) node [pos=1,below]{$j$};
\draw[thick](-.85,0) -- +(180-40:.5) node [pos=1,above]{$i-1$};
\draw[thick](-.85,0) -- +(180+40:.5) node [pos=1,below]{$j+1$};

\draw[thick](-.85,0) -- (.85,0);
\draw[fill=white](0,0) circle (.35cm);
\draw[fill=gray] (-.85,0) circle (0.35cm);
\draw[fill=gray] (.85,0) circle (0.35cm);

\end{scope}
\end{scope}

\begin{scope}[xshift=.5cm,yshift=+.5cm]
\begin{scope}[xscale=0.25,yscale=0.25]

\foreach \t in {-20,20}{
\draw[thick](.85,0) +(\t:.5) circle (1pt) ;
}
\foreach \t in {180-20,180+20}{
\draw[thick](-.85,0) +(\t:.5) circle (1pt) ;
}

\draw[thick](.85,0) -- +(40:.5) node [pos=1,above]{$i$};
\draw[thick](.85,0) -- +(-40:.5) node [pos=1,below]{$j$};
\draw[thick](-.85,0) -- +(180-40:.5) node [pos=1,above]{$i-1$};
\draw[thick](-.85,0) -- +(180+40:.5) node [pos=1,below]{$j+1$};

\draw[thick](-.85,0) -- (.85,0);
\draw[thick](0,.5) -- (0,0);
\draw[fill=white](0,.5) circle (.25cm);
\draw[fill=gray] (-.85,0) circle (0.35cm);
\draw[fill=gray] (.85,0) circle (0.35cm);

\end{scope}
\end{scope}
\end{tikzpicture}

\label{tadpolebubble}

\end{centering}
\end{figure}

\noindent Note that, despite its name, the adjacency swaps a \emph{cut} propagator $X_j$ with a tadpole propagator $X_{(i,\dots,i-1)}$. This time we have to focus on the two shared propagators at the sides of the bubble, whose variables are
\begin{align*}
dX_{(j+1,\dots,i-1)} &= -d(X_{j+1} + \dots + X_{i-2}) \\
dX_{(i,\dots,j)} &= -d(X_{i} + \dots + X_{j-1}), 
\end{align*}
we have
\begin{align*}
dX_{(i,\dots,i-1)} &= - d(\overline{ X_{i} + \dots + X_{j-1}} + X_j + \overline{X_{j+1} + \dots + X_{i-1}}) \\
&= -dX_j,
\end{align*}
where in the last passage we used the fact that the overlined terms vanish under wedge with the two shared propagators.

\noindent
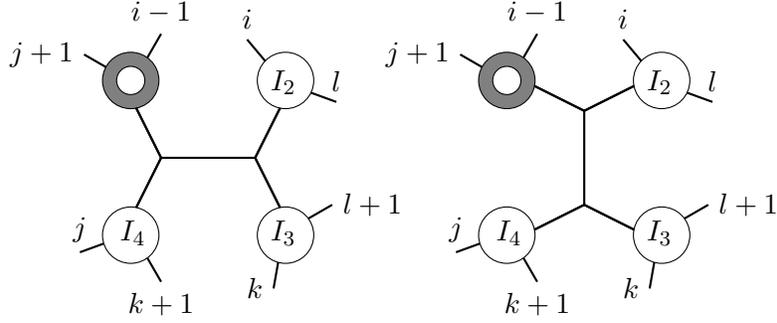
\emph{s-channel/u-channel adjacency}. Finally, we have a adjacency involving the tree structure of the diagram. In the figure we draw the loop part on the leg $I_1$ but its actual position is irrelevant, only that it is the same in both diagrams.

\FloatBarrier
\begin{figure}
\begin{centering}

\begin{tikzpicture}[>=latex,scale=5][every node/.style={transform shape}]

\begin{scope}[xshift=-.5cm,yshift=+.5cm]
\begin{scope}[scale=0.25]

\draw[thick] (0.8,0.8) -- ++(130:.6) node [pos=1,above]{$i$};
\draw[thick] (0.8,0.8) -- ++(-20:.6) node [pos=1,above]{$l$};

\draw[thick] (-0.8,0.8) -- ++(150:.6) node [pos=1,left]{$j+1$};
\draw[thick] (-0.8,0.8) -- ++(60:.6) node [pos=1,above]{$i-1$};

\draw[thick] (0.8,-0.8) -- ++(260:.6) node [pos=1,left]{$k$};
\draw[thick] (0.8,-0.8) -- ++(30:.6) node [pos=1,right]{$l+1$};
\draw[thick] (-0.8,-0.8) -- ++(200:.6) node [pos=1,above]{$j$};
\draw[thick] (-0.8,-0.8) -- ++(300:.6) node [pos=1,below]{$k+1$};

\draw[thick] (-.5,0) -- (.5,0);
\draw[thick] (-1,1) -- (-.5,0) -- (-1,-1);
\draw[thick] (-1,1) -- (-.5,0) -- (-1,-1);
\draw[thick] (1,1) -- (.5,0) -- (1,-1);
\draw[thick] (1,1) -- (.5,0) -- (1,-1);

\draw[fill=gray](-1,1)+(-45:0.25) circle (.3cm);
\draw[fill=white](-1,-1)+(45:0.25) circle (.3cm);
\draw[fill=white](1,-1)+(135:0.25) circle (.3cm);
\draw[fill=white](1,1)+(-135:0.25) circle (.3cm);

\draw[fill=white](-1,1)+(-45:0.25) circle (.15cm);

\node at (0.8,0.8) {$I_2$};
\node at (0.8,-0.8) {$I_3$};
\node at (-0.8,-0.8) {$I_4$};

\end{scope}
\end{scope}

\begin{scope}[xshift=.5cm,yshift=+.5cm]
\begin{scope}[xscale=0.25,yscale=0.25]

\draw[thick] (0.8,0.8) -- ++(130:.6) node [pos=1,above]{$i$};
\draw[thick] (0.8,0.8) -- ++(-20:.6) node [pos=1,above]{$l$};
\draw[thick] (0.8,-0.8) -- ++(30:.6) node [pos=1,right]{$l+1$};

\draw[thick] (-0.8,0.8) -- ++(150:.6) node [pos=1,left]{$j+1$};
\draw[thick] (-0.8,0.8) -- ++(60:.6) node [pos=1,above]{$i-1$};

\draw[thick] (0.8,-0.8) -- ++(260:.6) node [pos=1,left]{$k$};
\draw[thick] (-0.8,-0.8) -- ++(300:.6) node [pos=1,below]{$k+1$};

\draw[thick] (-0.8,-0.8) -- ++(200:.6) node [pos=1,above]{$j$};

\begin{scope}[rotate=90]
\draw[thick] (-.5,0) -- (.5,0);

\draw[thick] (-1,1) -- (-.5,0) -- (-1,-1);
\draw[thick] (-1,1) -- (-.5,0) -- (-1,-1);
\draw[thick] (1,1) -- (.5,0) -- (1,-1);
\draw[thick] (1,1) -- (.5,0) -- (1,-1);

\draw[fill=white](-1,1)+(-45:0.25) circle (.3cm);
\draw[fill=white](-1,-1)+(45:0.25) circle (.3cm);
\draw[fill=white](1,-1)+(135:0.25) circle (.3cm);
\draw[fill=gray](1,1)+(-135:0.25) circle (.3cm);
\end{scope}
\draw[fill=white](-1,1)+(-45:0.25) circle (.15cm);
\node at (0.8,0.8) {$I_2$};
\node at (0.8,-0.8) {$I_3$};
\node at (-0.8,-0.8) {$I_4$};

\end{scope}
\end{scope}
\end{tikzpicture}

\label{suchannel}
\end{centering}
\end{figure}

\noindent
All the four shared variables $X_{I_j}$ have to be kept in consideration, and in particular remember that $dX_{I_1} = -d(X_i + \dots + X_{j-1})$ \footnote{We made a slight abuse of notation, as we should write $X_{(I_1)^c}$ rather than $X_{I_1}$}.
For the diagram on the left we have
\begin{align*}
&\dots \wedge d X_{I_1} \wedge dX_{I_4} \wedge dX_{I_2 I_3} \\
= &\dots \wedge d X_{I_1} \wedge dX_{I_4} \wedge d(\overline{X_{I_2 I_3} + X_{I_4} - X_k} + X_k) \\
= &\dots \wedge d X_{I_1} \wedge dX_{I_4} \wedge dX_k, 
\end{align*}
we freely added a shared propagator and recognised the overlined term as $dX_{I_1}$.

\noindent
Similarly, for the other diagram we get 
\begin{align*}
\dots \wedge dX_{I_3 I_4} = \dots \wedge dX_{l},
\end{align*}

\noindent
finally we note that 
\begin{align*}
\dots \wedge d(X_k + X_{l}) = \dots \wedge d(X_k + X_{l} + X_{I_2} + X_{I_3} + X_{I_4}) = \dots \wedge d(X_{I_1}) = 0,
\end{align*}
thus proving \eqref{eq:mutationdx} also in this case. 

By virtue of \eqref{eq:mutationdx}, we can again write the canonical form \eqref{eq:canonical} using a single measure, we choose $d^n X = \bigwedge\limits_{i=1}^n dX_i$. Doing so we obtain
\begin{align}
\label{eq:almost}
\Omega_{\mathrm{H}_n} = \underline{\Omega}_{\mathrm{H}_n} d^n X,
\end{align}
where $\underline{\Omega}_{\mathrm{H}_n}$ is given by the sum over 1-loop planar diagrams. In the sum are involved also UV/IR tadpoles and diagrams with bubbles on external legs. Such unphysical contributions appear with terms $X_{0},X_{i,\dots,i+n}$ or $X_{i,\dots,i+n-1}$ in the denominator, which in turn are given by expressions linear in the $X_i$ and in the various $\epsilon_I$. Therefore, we can kill the external bubbles and the tadpoles by taking the limit $\epsilon_I \to \infty$ for those $I$ which correspond to tadpole and bubble facets.
After that, if we substitute $X_I \to s_I$, $s_I$ being the physical propagator associated to $X_I$, we are left with the 1-loop integrand!
More precisely, note that each variable $X_I$ carries an $\epsilon_I$ term uniquely associated to it. Therefore, we can first solve the $\epsilon_I$ for \emph{all} the $X$'s and then the substitution $X_I \to s_I$ can be done unambiguously, even if we have an expression for $\underline{\Omega}_{\mathrm{H}_n}$ where the constants $\epsilon_I$ and the variables $X_i$ are not manifestly appearing in a combination from which we can recognise a variable $X_I$. 

Before ending this section, we would like to give another interpretation of the $\epsilon \to \infty$ limit. In order to do so, it is convenient to switch to a projective language.
We think of the coordinates $X_i$ of our abstract space as affine coordinates on $\mathbb{P}^n$, i.e. we introduce the projective vector
$$Y = (1, X_1 , \dots , X_n).$$
Facets are given by linear equations of the form
\begin{align}
\label{eq:YWF}
X_f = Y \cdot W_f = 0,
\end{align}
for a suitable dual vector $W_f$, which is again naturally projective.
For example, for $n=4$, the UV facet has a dual vector given by
$$W_{\mathrm{UV}} = (\epsilon_0 - \sum_{i=1}^4 \epsilon_{(i,i+1)}, 1, 1, 1, 1),$$
and by taking the projective limit this becomes
$$W_{\mathrm{UV}} = (1,0,0,0,0),$$
looking back at \eqref{eq:YWF} it is now clear that we need $Y=(0,*,*,*,*)$, that is the facet has moved to the hyperspace at infinity.
The $\epsilon \to \infty$ prescription has then a simple projective meaning: it deforms the Halohedron so that its un-physical facets are at infinity and possibly degenerate depending on the ratios of the $\epsilon$ that are sent to infinity.

\section{Recursion Formula for the 1-loop Integrand}
\label{Recursion Formula for the 1-loop Integrand}

We understood how to extract the 1-loop integrand from the canonical form of the Halohedron, reproducing the Feynman diagram representation. However, we can obtain new formulae by considering triangulations of the Halohedron to compute its canonical form. In this section we provide an example of this, using a recursion formula for the canonical form of a polytope that works by recycling the canonical forms of its facets. We just sketch this construction here, further details can be found in full generality in \cite{Arkani-Hamed:2017mur}. The idea is to triangulate the polytope using a reference point $Z^*$ in its interior. For each facet $\mathcal{F}$ we take the convex hull $\mathcal{P}_\mathcal{F}$ of its vertices with $Z^*$, together these polytopes triangulate $\mathcal{P}$. The canonical form of $\mathcal{P}_\mathcal{F}$ can be obtained by a suitable deformation of the canonical form of the facet $\mathcal{F}$. In the case of the Associahedron and the Halohedron, these canonical forms can be interpreted as lower points amplitudes, therefore we obtain a geometrically inspired recursion formula.
\indent We will explicitly unwind this procedure for $n=4$. We choose as reference point the intersection of the cut facets $X^* = (0,0,0,0)$. Therefore, all facets will contribute to the recursion, except for the cut facets for which the convex hull with $X^*$ is $2$ dimensional rather than $3$ dimensional. The recursion, written directly at the level of the rational functions, reads
\begin{align}
\label{eq:4recursion}
\underline{\Omega}_{\mathrm{H}_4} = \sum_{I \ne (i)} \hat{\underline{\Omega}}_{X_I},
\end{align}
where the sum runs over all facets of $\mathrm{H}_4$, excluding the cut facets, and $\hat{\underline{\Omega}}_{X_I}$ is a deformation of the rational function of the facet $X_I$ obtained as follows. First, we write $X_I = X_I^0 + X_I^\prime$, where in $X_I^0$ we gather the constants $\epsilon$ and in $X_I^\prime$ the basis variables $X_i$. The deformation is then defined by replacing each basis variable appearing in $\underline{\Omega}_{X_I}$ with
\begin{align*}
X_i \to \hat{X}_i = -\frac{X_I^0}{X_I^{'}} X_i,
\end{align*}
and multiplying by the overall factor $(X_I^0/X_I^\prime)^4$. In other words we have that
\begin{align}
\label{eq:canonicalfacet}
\hat{\underline{\Omega}}_{X_I}(\epsilon| X_i) = \left(-\frac{X_I^0}{X_I^{'}}\right)^4 \frac{\underline{\Omega}_{X_I}(\epsilon|-\frac{X_I^0}{X_I^{'}} X_i)}{X_I},
\end{align}
where we emphasized that the rational function $\underline{\Omega}_{X_I}$ depends on both the constants $\epsilon$ and the basis variables $X_i$. We are interested in the limits $\epsilon_J \to \infty$, for $J=0$ or $|J|=4,3$, and which we take in this same order. Recalling the Feynman diagram representation for the rational functions $\underline{\Omega}_{X_I}$, it is easy to understand the effect of this limit on \eqref{eq:canonicalfacet}. If $X_I$ is not one of the facet involved in the limit, i.e. it has $|I| \le 2$, then tadpoles and external bubbles do not contribute to its rational function. Also, because of the order of the limits, UV tadpoles do not appear in the rational function of an IR tadpole facet, but internal bubbles do.
Finally, if $I$ is one of the facet being sent to infinity one has to take into account the diverging pre-factor $X_I^0$ in \eqref{eq:canonicalfacet}. A quick power counting shows that the net effect is that we are left with
\begin{align}
\hat{\underline{\Omega}}_{X_I}(\epsilon| X_i)= \frac{\underline{\Omega}_{X_I}(0| X_i)}{X_I^\prime}.
\label{eq:limitcanonical}
\end{align}
i.e. we forget the dependence on the constants $\epsilon$ of the propagators appearing in $\underline{\Omega}_{X_I}/X_I$.

We now show explicitly the computation for each facet contributing to \eqref{eq:4recursion}.

\noindent
\emph{UV facet}. This facet is associated to the variable $$X_0 = \sum_{i=1}^4 \epsilon_i - \epsilon_0 - \sum_{i=1}^4 X_i,$$ therefore $$X_0^0 = \sum_{i=1}^4 \epsilon_{i} - \epsilon_0 \quad \quad X_0^{'} = -\sum_{i=1}^4 X_i.$$
The rational function $\underline{\Omega}_{X_0}$ is given by a sum over all 20 UV-tadpole diagrams. If we group those associated with the same IR tadpole propagator, say $X_{(i,i+1,i+2,i+4)}$, we obtain
\begin{align}
\label{eq:5ptcontr}
&\frac{1}{X_{(i,i+1,i+2,i+4)}}\left( \frac{1}{X_{(i,i+1)} X_{(i+2,i+3)}} + \frac{1}{X_{(i+1,i+2)} X_{(i+1,i+2,i+3)}}\right. + \nonumber \\
&\left. \frac{1}{X_{(i,i+1)} X_{(i,i+1,i+2)}} + \frac{1}{X_{(i+1,i+2,i+3)} X_{(i+2,i+3)}} + \frac{1}{X_{(i,i+1,i+2)} X_{(i+1,i+2)}} \right).
\end{align}
Since this facet is going to infinity, we forget the $\epsilon$ dependence of the planar variables appearing in \eqref{eq:5ptcontr}, after which the terms in the bracket of \eqref{eq:5ptcontr} sum up to 
$$\frac{X_i + X_{i+1} + X_{i+2}}{X_i X_{i+1} X_{i+2}},$$
the numerator cancels with the denominator outside the bracket leaving us with 
$$\frac{1}{X_i X_{i+1} X_{i+2}},$$
and summing over the four IR tadpole propagators we get
\begin{align}
\label{eq:uvrational}
\underline{\Omega}_{X_0}(0|X_i) = \frac{X_1 + X_{2} + X_{3} + X_{4}}{X_1 X_{2} X_{3} X_{4}},
\end{align}
plugging \eqref{eq:uvrational} back in \eqref{eq:limitcanonical} we finally get the contribution of the UV facet to the recursion
\begin{align}
\label{eq:UVcontribution}
\hat{\underline{\Omega}}_{X_0} = \pm \frac{1}{X_1 X_2 X_3 X_4}.
\end{align}

\noindent
\emph{Tadpole facet}. We focus on the facet $X_{(1,2,3,4)}$, the remaining ones are obtained through a cyclic shift.
The rational function $\underline{\Omega}_{X_{(1,2,3,4)}}$ is given by a sum over 10 IR and UV tadpole diagrams. However, the UV are killed by the limit $\epsilon_0 \to \infty$ which we take before the limit $\epsilon_{(1,2,3,4)} \to \infty$. Therefore we are left with 5 IR tadpole diagrams, whose contribution is almost identical to \eqref{eq:5ptcontr} a part for the prefactor:
\begin{align}
&\frac{1}{X_{4}}\left( \frac{1}{X_{(1,2)} X_{(3,4)}} + \frac{1}{X_{(2,3)} X_{(2,3,4)}}\right. + \nonumber \\
&\left. \frac{1}{X_{(1,2)} X_{(1,2,3)}} + \frac{1}{X_{(2,3,4)} X_{(3,4)}} + \frac{1}{X_{(1,2,3)} X_{(2,3)}} \right),
\label{eq:tadpole5ptcontr}
\end{align}
again because of the limit the variables in \eqref{eq:tadpole5ptcontr} lose the $\epsilon$ dependence so that $X_{(i,i+1)} \to - X_i$ and $X_{(i,i+1,i+2)} \to - X_i - X_{i+1}$, after some manipulations we get
$$\underline{\Omega}_{X_{(1,2,3,4)}}(0|X_i) = \frac{X_1 + X_{2} + X_{3}}{X_1 X_{2} X_{3} X_{4}},$$
and then
\begin{align}
\label{eq:tadpolecontribution}
\hat{\underline{\Omega}}_{X_{(1,2,3,4)}} = \pm \frac{1}{X_1 X_2 X_3 X_4}.
\end{align}

\noindent
\emph{Bubble facet}. Again we consider a specific case, for instance $X_{(1,2,3)} = 0$. Recall that this facet factorises in $\rm{H}_\mathcal{A} \times \mathcal{A}_1$, $\mathcal{A}_1$ being a 4-point Associahedron. Accordingly, we can write its rational function as
\begin{align*}
\underline{\Omega}_{X_{(1,2,3)}} = \left(\frac{1}{X_{(1,2)}} + \frac{1}{X_{(2,3)}} \right) \times \left(\frac{1}{X_3 X_4} + \dots \right),
\end{align*}
where in $\dots$ we gathered terms that vanish due to the limits $\epsilon_0, \epsilon_I \to \infty$ with $|I|=4$, thus in the limit we have
\begin{align*}
\underline{\Omega}_{X_{(1,2,3)}}(0|X_i) = \frac{X_1+X_2}{X_1 X_2 X_3 X_4},
\end{align*}
and finally once again
\begin{align}
\label{eq:bubblecontribution}
\hat{\underline{\Omega}}_{X_{(1,2,3)}} = \pm \frac{1}{X_1 X_2 X_3 X_4}.
\end{align}

\noindent
\emph{Factorisation Facet}. Consider $X_{(1,2)} = \epsilon_1 - X_1 = 0$, there are only two diagrams surviving all the limits so the rational function is given by
\begin{align*}
\underline{\Omega}_{X_{(1,2)}} &= \left(\frac{1}{X_2 X_3 X_4} + \frac{1}{X_2 X_4 X_{(3,4)}}\right) 
\end{align*}
which plugging into \eqref{eq:canonicalfacet} gives
\begin{align}
\label{eq:factorisationcontribution}
\hat{\underline{\Omega}}_{X_{(1,2)}} = \frac{\epsilon_1}{X_1 X_2 X_4 (\epsilon_1 - X_1)} \left(\frac{1}{X_3} +\frac{\epsilon_1}{X_1\epsilon_3 - \epsilon_1 X_3}\right).
\end{align}
Note that the denominator $X_1\epsilon_3 - \epsilon_1 X_3$ represents a \emph{spurious} pole, it eventually cancels with a similar contribution coming from the facet $X_{(3,4)} = 0$.

Putting the contributions from the various facets together, and paying attentions to the signs involved, we find that each tadpole contribution cancels with a corresponding bubble. We are left with the UV contribution and the factorisation contributions. 

Next, we need to translate back \eqref{eq:factorisationcontribution} into usual loop and external variable. First have to rewrite the constants $\epsilon_i$ in terms of the planar variables $X_I$ and then substitute the corresponding propagators:
$$\epsilon_i = X_{(i,i+1)} + X_{i} \to s_{i,i+1} + \ell_i^2,$$
where $s_{i,j} = 2 k_i \cdot k_j$ and $\ell^\mu_i$ is the momentum flowing between particles $i$ and $i+1$, e.g. $\ell^\mu_1 = \ell^\mu$, $\ell^\mu_2 = \ell^\mu + k^\mu_2$ and so on. After this is done, with a bit of algebra we get the following expression for the 4-point integrand
\begin{align}
\label{eq:recursedamplitude}
m^{\rm{1-loop}}_{4} &=  -\frac{1}{\ell_1^2 \ell_2^2 \ell_3^2 \ell_4^2} + \left(\frac{(\ell_1^2 + s_{12}) (\ell_3^2 + s_{12}  )}{s^2_{12} \ell_2^2 \ell_3^2  \ell_4^2  (\ell_3^2 - \ell_1^2)} + \mathrm{cyclical}\right),
\end{align}
where the sum is over the remaining three cyclically shifted terms. Note that \eqref{eq:recursedamplitude} has double poles $s_{12} = 0$ and $s_{23} = 0$ coming from the internal bubbles. If we expand around $s_{12} = 0$ we get
$$\frac{1}{s_{12}^2} \frac{1}{\ell_2^2\ell_4^2} + \frac{1}{s_{12}} \left(\frac{1}{\ell_1^2 \ell_2^2 \ell_4^2} + \frac{1}{\ell_2^2 \ell_3^2 \ell_4^2} \right) + \dots $$
and from the coefficients of the expansion we read the internal bubble contribution and the \emph{two} contributions to the residue $s_{12} = s_{34}= 0$. 

\section{Conclusions and Outlook}

In this paper we have shown that it is possible to extract the 1-loop amplitude from the canonical form of an Halohedron realised as a convex polytope in an abstract kinematical space, thus proving the conjecture we made in our previous work: The Halohedron is the 1-loop Amplituhedron. The fact that an integrand with double poles can be obtained from the canonical form of a positive geometry is a remarkable proof of principle\footnote{During the preparation of this manuscript, we realised that this possibility was already been hinted at in \cite{Arkani-Hamed:2017mur} based on the heuristic that non-logarithmic theories may be obtained as effective theories of logarithmic ones.}, and gives new strength to the idea that some deep physical concept, lurking in the shadows of the Lagrangian formalism of Quantum Field Theories, is captured by positive geometries.

There are many directions for future investigations. The most natural one is to move at higher loop level, considering the moduli spaces of the Poincare' disk with several circles evicted, using the notation of \cite{devadoss}, this is $\mathcal{M}_{(0,\ell+1)(0,n)}$. However, as proven there, it fails to be a polytope. The reason is the presence of a geodesical arc whose contraction lowers the dimension by two. Nevertheless, this arc is equivalent to the UV arc of the Halohedron, and it is not associated to a physical singularity of the integrand. Therefore, it is likely possible to hide this problem at infinity, following the same spirit we did here for the UV facet. 

A somewhat simpler generalisation of our work would be to find an expression for integrands with two different orderings. In light of the lessons from the tree level story \cite{Arkani-Hamed:2017mur,Mizera:2017cqs,Mizera:2017rqa,He:2018pue}, it is quite natural to expect them to be found from intersection of Halohedra sitting in the full moduli space $\mathcal{M}_{1,n}$, or by pull back of a single 1-loop planar scattering form to the intersection of two $n$-dimensional abstract spaces. 

A third interesting avenue would be to study other triangulations of the Halohedron and the associated recursion formulae for the integrand. For example, it would be interesting to try to reproduce the ring diagram equality - which underpins the forward limit formula of \cite{He:2015yua} - using as reference point for the triangulation $X^* = (\epsilon_1, \dots , \epsilon_n)$, so that the cut facets do count in the recursion formula. It would be also nice to understand if it is possible to give an interpretation of partial fractions identities, which can be obtained by residue theorems \cite{Geyer:2015jch}, using the geometry of the Halohedron.

Finally, we mentioned that the $\epsilon \to \infty$ limit has a natural projective meaning: it is actually sending the corresponding facets to hyperplanes at infinity. Therefore, it is tempting to define a limit positive geometry whose canonical form directly gives the integrand. It would be fascinating to understand if this is possible, and probably would give a more beautiful and geometrical understanding on the way tadpoles and bubbles cancel each other.

\section{Acknowledgments}

I would like to thank S. Cacciatori and G. Ruzza for valuable conversations.

\end{document}